\documentclass[useAMS,usenatbib]{mn2e}

\usepackage{verbatim}
\usepackage{graphicx}
\usepackage{amssymb}
\usepackage{amsmath}
\usepackage[figuresright]{rotating}
\newcommand{\mnras}{MNRAS}

\newcommand{\aap}{A\&A}
\newcommand{\prd}{PhRvD}
\newcommand{\apj}{ApJ}
\newcommand{\aj}{AJ}

\title[Can Early Dark Energy be Detected in Non-Linear Structure?]{Can
Early      Dark     Energy      be     Detected      in     Non-Linear
Structure?\footnotemark[1]} \author[Francis, Lewis \& Linder] {Matthew
J.  Francis$^{1}$\thanks{Email: mfrancis@physics.usyd.edu.au}, Geraint
F. Lewis$^{1}$ and  Eric V. Linder$^{2}$ \\ $^{1}$  School of Physics,
University  of  Sydney, NSW  2006,  Australia\\  $^{2}$ University  of
California, Berkeley Lab, Berkeley, CA 94720, USA }

\begin{document}

\date{}

\pagerange{\pageref{firstpage}--\pageref{lastpage}} \pubyear{2007}

\maketitle

\label{firstpage}

\begin{abstract}
We  present the  first study  of early  dark energy  cosmologies using
N-body  simulations   to  investigate  the   formation  of  non-linear
structure. In contrast  to expectations from semi-analytic approaches,
we find that early dark energy  does not imprint a unique signature on
the statistics of non-linear structures.  Investigating the non-linear
power spectra  and halo  mass functions, we  show that  universal mass
functions hold for early dark energy, making its presence difficult to
distinguish  from $\Lambda$CDM.   Since early  dark energy  biases the
baryon acoustic oscillation scale, the lack of discriminating power is
problematic.
\end{abstract}

\begin{keywords}
methods:N-body simulations --- methods: numerical --- dark matter --- 
dark energy --- large-scale structure of Universe
\end{keywords}

\long\def\symbolfootnote[#1]#2{\begingroup%
  \def\thefootnote{\fnsymbol{footnote}}\footnotetext[#1]{#2}\endgroup} 

\symbolfootnote[1]{Research
  undertaken  as part  of  the Commonwealth  Cosmology Initiative  (CCI:
  www.thecci.org),  an  international  collaboration  supported  by  the
  Australian Research Council}

\section{Introduction}

Uncovering the nature of the dominant energy component of the Universe
at  the  current epoch,  dark  energy, is  a  central  goal of  modern
precision cosmology. The evidence for  the existence of dark energy is
strong  (\citet{riess98}, \citet{perl99},  \citet{wmap5}),  however on
the theoretical front there are no leading candidates despite a wealth
of suggestions  (see \citet{review}  for a review).   As observational
data  becomes more  precise and  extensive due  to current  and future
generations  of large surveys  and increasingly  powerful instruments,
there  remains a number  of theoretical  challenges to  determine with
high accuracy the expected observational consequences of the myriad of
possible dark energy models.

Dark  energy  affects  the   Universe  primarily  through  the  global
expansion  rate.   This  can  be measured  directly  through  distance
measurements  such  as   from  Type  Ia  supernovae  (\citet{riess98},
\citet{perl99},  \citet{kowalski08}) over the  last 10  billion years.
However, some dark energy models predict a non-negligible contribution
to the early time expansion  behavior, called early dark energy (EDE).
To  probe EDE  requires observations  tied to  the very  high redshift
universe,  such as  the cosmic  microwave background  (CMB)  or baryon
acoustic    oscillations     (BAO).     However,    as     shown    in
\cite{LinderRobbers},  early   dark  energy  can  be   hidden  in  CMB
observations, even  while shifting  the intrinsic acoustic  scale upon
which BAO measurements rely upon.  Thus failure to detect EDE can bias
the cosmological model interpreted from CMB and BAO data.

An alternative and complementary measure of dark energy is through the
observation of structure in the  Universe. The change in the dynamical
evolution of expansion relative to the $\Lambda$CDM model, say, alters
the  growth  of  structures  in  the inhomogeneous  universe  that  we
inhabit.   The statistics of  structure therefore  encodes information
about dark energy. Structure measures involve a different weighting of
cosmological parameters  than distance measures,  thus the combination
breaks  degeneracies.   Moreover,  they  can  test  the  framework  of
gravitational growth of  inhomogeneities, important for distinguishing
between a  physical dark  energy and changes  to the laws  of gravity.
Finally,  growth is a  cumulative process,  depending on  the physical
conditions from early  times through the epoch at  which the structure
is  observed.    Thus,  observations  of  structure   are  crucial  to
understanding cosmology at all  epochs. Because of the implications of
early dark energy  for interpreting CMB and BAO  measurements, and for
understanding the fundamental physics behind dark energy, the behavior
of structure formation and evolution is a key point.

Early dark energy models address the coincidence problem by having the
dark energy  evolve relatively slowly compared to  the dominant matter
or radiation component; in some  cases motivated by string theory they
scale exactly with the background  component.  Indeed, this was one of
the first dark energy models \citep{Wett88}.  EDE is discussed in some
detail in  \cite{DoranRobbers06} and references  therein.  Constraints
from  CMB  and primordial  nucleosynthesis  data  restrict the  energy
fraction in the  early universe contributed by EDE to  less than a few
percent  \citep{DoranRW07} but  this could  have  significant effects.
Some  of   the  implications  for  linear  growth   were  examined  in
\cite{Linder06}     and     \citet{DoranRobbers06}    and     extended
semi-analytically to  non-linear theory by  \cite{bartDW06}.  The main
result of previous studies of the non-linear growth in EDE cosmologies
(such  as \citet{bartDW06}  and  \citet{FedeliBart07}) is  that for  a
fixed   linear  theory   matter  perturbation   amplitude   at  $z=0$,
$\sigma_8$,  there are  more collapsed  structure in  EDE  models than
$\Lambda$CDM.   For  higher   redshifts  this  effect  increases,  for
instance \cite{bartDW06} found that at $z=1$, there are up to an order
of magnitude more dark matter  halos for higher halo masses ($M_{halo}
\gtrsim 10^{15} M_{\odot}$).  This  can be simply interpreted as early
dark  energy,  which does  not  appreciably  cluster, suppressing  the
growth of  structure at early  times.  Hence, to achieve  the observed
level  of current structure,  the early  amplitude of  clustering must
have   been   greater.     Other   applications   include   the   high
Sunyaev-Zel'dovich amplitude  possibly seen at high  multipoles in the
CMB \citep{SadehRephSilk,WaizmanBart}  and the early  formation of the
first stars \citep{SadehReph08}.

A key aim of this paper  is to use N-body simulations to calculate the
predicted number of dark matter halos in EDE cosmologies compared with
$\Lambda$CDM.   In particular,  the non-linear  level of  structure is
affected  by the early  growth in  a way  that can  change predictions
based on  the usual  linear growth factor  at a particular  epoch.  We
examine both the halo mass  function giving the abundance of collapsed
objects and the non-linear  matter power spectrum.  N-body simulations
are a  robust method  of determining these  statistics and  this paper
critically examines predictions previously made analytically.

In Section \ref{simdetail} we  describe the N-body simulations we used
to  probe the  effects of  EDE on  non-linear structure  formation. We
present the results for the  abundance of dark matter halos in Section
\ref{halos} and  the results for the non-linear  matter power spectrum
in Section \ref{psresults}. We  discuss the implication of our results
and make concluding remarks in Section \ref{conclude}.

\section{Simulation Details}\label{simdetail}

The  simulations were  performed  using the  cosmological N-body  code
GADGET2 \citep{Gadget  code} suitably modified  in order to  model the
expansion history  for arbitrary dark energy models.   The initial CDM
power   spectrum   was  determined   using   the   CMB  code   CMBEASY
\citep{2005JCAP...10..011D}.   Initial  realisations  of  the  density
field were created using part  of the COSMICS code \citep{COSMICS code
paper}, modified  to allow an arbitrary initial  power spectrum. Given
the  real space  displacement  field, $\Psi_0(\textbf{x})$,  generated
from the  linear power spectrum  $P(k)$ at redshift zero,  the initial
grid  of particles  are perturbed  using the  Zel'dovich approximation
\citep{zed} resulting in the  positions and velocities at the starting
scale factor $a_i$ of

\begin{equation}
\textbf{x} = \textbf{q} + \frac{D(a_i)}{D(1)}\Psi_0(\textbf{q})
\end{equation}
\begin{equation}
\dot{\textbf{x}} = \frac{\dot{D}(a_i)}{D(1)}\Psi_0(\textbf{q})
\end{equation}
where $\textbf{q}$ are the real  space grid co-ordinates and $D(a)$ is
the linear growth  factor. The dots refer to  derivatives with respect
to cosmic time  $t$. The linear growth factor  and its derivative were
calculated by numerically integrating  the growth equation detailed in
\citet{LinderJenkins}.  In  all cases, the starting  redshift was kept
consistent  between the  different  physical models.   In general  the
linear power  spectrum shape varies  in the different  models compared
and therefore  we made initial  conditions separately for  each model,
always ensuring  that the  initial fluctuation amplitude  returned the
desired  $\sigma_{8}$  at  $z=0$.   The power  spectra  measured  from
simulation outputs were  compared to linear theory to  ensure that the
large scale power evolved  as expected.  All simulations and numerical
analysis were  performed on `The Green  Machine' supercomputer located
at   the  Centre  for   Astrophysics  and   Supercomputing,  Swinburne
University.

The early dark energy  model is parametrised through its dimensionless
energy  density as  a  function of  redshift by  \citep{DoranRobbers06}
\begin{equation}\label{edeparam}
\Omega_d(a)              =              \frac{\Omega_d^0             -
\Omega_e(1-a^{-3w_0})}{\Omega_d^0+\Omega_m^0a^{3w_0}}               +
\Omega_e(1-a^{-3w_0})
\end{equation}
in which  $\Omega_m^0$ and $\Omega_d^0$ are the  density parameters of
matter  and dark  energy  today ($\Omega_m^0=1-\Omega_d^0$  in a  flat
universe  as assumed here),  $w_0$ is  the equation  of state  of dark
energy today  and $\Omega_e$  is the energy  density fraction  of dark
energy at early times, $a\ll  1$. Note that our $\Omega_e$ corresponds
to $\Omega_d^e$ from  \cite{DoranRobbers06}.  Strictly for the initial
comparison  to  \cite{bartDW06}, we  also  use  the  bending model  of
\cite{Wetterich04}, defined as
\begin{equation}\label{wetEDE}
\ln\frac{\Omega_d(a)}{\Omega_m(a)}=\ln\frac{\Omega_d^0}{\Omega_m^0}-
\frac{3w_0\ln a}{1-b\ln a}. 
\end{equation}  
In  both cases  the dark  energy  equation of  state is  given by  the
standard formula $w(a)=-(1/[3\Omega_d (1-\Omega_d)])d\Omega_d/d\ln a$.
Note that \cite{DoranRW07} find that the marginal distributions of the
likelihood  for cosmological  parameters  other than  dark energy  are
insensitive to the specific parametrisation.

In all  the simulations using the  EDE form of  Eq. \ref{edeparam} the
cosmological  model used  was  a flat  universe with:  $\Omega_m=0.3,$
$\Omega_b=0.048,$  $h=0.69,$ $n=0.98,$ $\sigma_8=0.76,$  and $w_0=-1$,
except where specified otherwise.

Dark matter halos in our  simulation outputs were determined using two
different methods.  The  first is the Friends of  Friends (FOF) method
\citep{FOFmethod}  using  a   constant  linking  length  parameter  of
$b=0.2$.   An alternative  approach  to defining  halos in  simulation
outputs is  the spherical  overdensity method (SO)  (see \citet{lukic}
for a detailed comparison of FOF  and SO halo methods). To find the SO
defined   halos,  we   employed  the   \textsc{MPI}  version   of  the
\textsc{AMIGA} Halo Finder
\footnote{\textsc{AMIGA}   is  freely   available   for  download   at
http://www.aip.de/People/AKnebe/AMIGA/}  (\textsc{AHF}),  successor of
\textsc{MHF} introduced by  \citet{MHF}. In order to use  this for EDE
cosmologies,  we  made the  appropriate  modifications  to the  virial
overdensity calculation.

The matter  power spectrum was calculated utilising  the `chaining the
power' method, as  described in \citet{Halofit}, using the  cloud in a
cell grid assignment scheme.

The bulk of the simulations  used $256^3$ particles in a $256$ Mpc/$h$
box starting  at redshift $z=24$.   For each model we  performed eight
simulations  using  different  realisations  of  the  initial  density
field. The power  spectrum and halo mass functions  were averaged over
all realisations.   Additional simulations were  performed halving and
doubling  the box  size as  well  as altering  the starting  redshift,
softening  length and  accuracy parameters  to ensure  convergence. We
also check  that for $\Omega_e$  approaching zero, the  power spectrum
and  halo results  converged to  the results  for constant  $w=w_0$ as
expected. The results of these tests indicated that the power spectrum
ratio results  are at most  $\sim 1\%$ altered by  different numerical
parameters,  with  the most  important  factor  being  box size  which
affected  the power  spectrum  ratios mainly  at  high $k$.   Starting
redshift and  softening length make  little difference to  the ratios.
The halo mass function results  were somewhat affected by changing box
size, our  results agreed with the analysis  of \citet{PowerKnebe} for
the  effect of  box  size on  the  mass function.   The ratio  results
fluctuated at  the level of $\sim  10\%$ with varying box  size. We do
not claim  state of  the art precision  in our mass  function results,
however they  are sufficiently accurate  and converged to  support our
main thesis, as detailed in Section \ref{halos}.

\section{Halo Mass Function}\label{halos}

The halo mass  function (hereafter HMF), defined simply  as the number
of halos  of a  given mass  expected per unit  volume, $n(M,z)$,  is an
important theoretical  and observational statistic of  the dark matter
density  field.   Indeed,  the   abundance  is  often  claimed  to  be
exponentially  sensitive to  the  cosmology.  The  pioneering work  of
\citet{PressSch} related  the expected mass function  to the variance,
$\sigma(R)$ on some  length scale $R$ of the  linearly evolved density
field and the linear  collapse parameter, $\delta_c$, which is derived
from spherical collapse arguments and  for $\Lambda$CDM is only a weak
function of cosmology and redshift.

The advent of larger N-body  simulations allowed this mass function to
be  tested rigorously,  resulting  in improvements  that modified  the
basic functional form to  include ellipsoidal collapse with parameters
fit to  simulation by \citet{ShethTormen}.  This in  turn was improved
upon by \citet{Jenkins01} and  more recently by \citet{Warren} with an
alternative approach of fitting directly the multiplicity function
\begin{equation}\label{multipfunc}
f(\sigma)=\frac{M}{\rho}\frac{dn}{d\textrm{ln}\sigma^{-1}}
\end{equation}
by some  function of  $\sigma$  without reference  to the  collapse
parameter.   Here $\rho$ is  the background  matter density.   In both
\citet{Jenkins01}  and  \citet{Warren}   universal  formulas  for  the
multiplicity function,  valid for a range of  cosmologies, were found.
The  effect of  cosmology enters  through the  variance $\sigma$  as a
function of scale  and redshift and the mean  matter density.  Writing
the Press-Schechter  style mass functions  in this form  gives results
that are  broadly in  agreement with the  Jenkins and  Warren formulas
with  the latter  being  more  accurate fits  to  numerical data,  but
qualitatively  similar.  Despite  being formulated  from $\Lambda$CDM,
matter dominated  and open cosmologies  only, the Jenkins  formula has
been  shown to  be  valid also  for  a large  N-body  simulation of  a
dynamical dark energy cosmology \citep{LinderJenkins}.

In previous work on EDE  cosmologies relying on estimation of the HMF,
(such as  \citet{bartDW06} and \citet{FedeliBart07})  the Sheth-Tormen
mass function has been used  to predict the relative halo abundance of
EDE and $\Lambda$CDM cosmologies.   This required determination of the
linear collapse parameter for  EDE cosmologies, for the derivation see
\citet{bartDW06}.  The  analysis using this mass  function suggested a
significant  increase in  the number  of massive  halos in  EDE models
compared to $\Lambda$CDM models  with the same $\sigma_8$ today.  This
was particularly  true at  $z=1$.  However, it  is not clear  that the
Press-Schechter  style   mass  functions,  relying   on  the  collapse
parameter, is to be preferred for these cosmologies over the universal
mass function  approach of Jenkins  and of Warren.   For $\Lambda$CDM,
the difference between the approaches is solely the degree of accuracy
across a range of halo masses, but the functions broadly agree.

\begin{figure}
\includegraphics[
  scale=0.35,
  angle=-90]{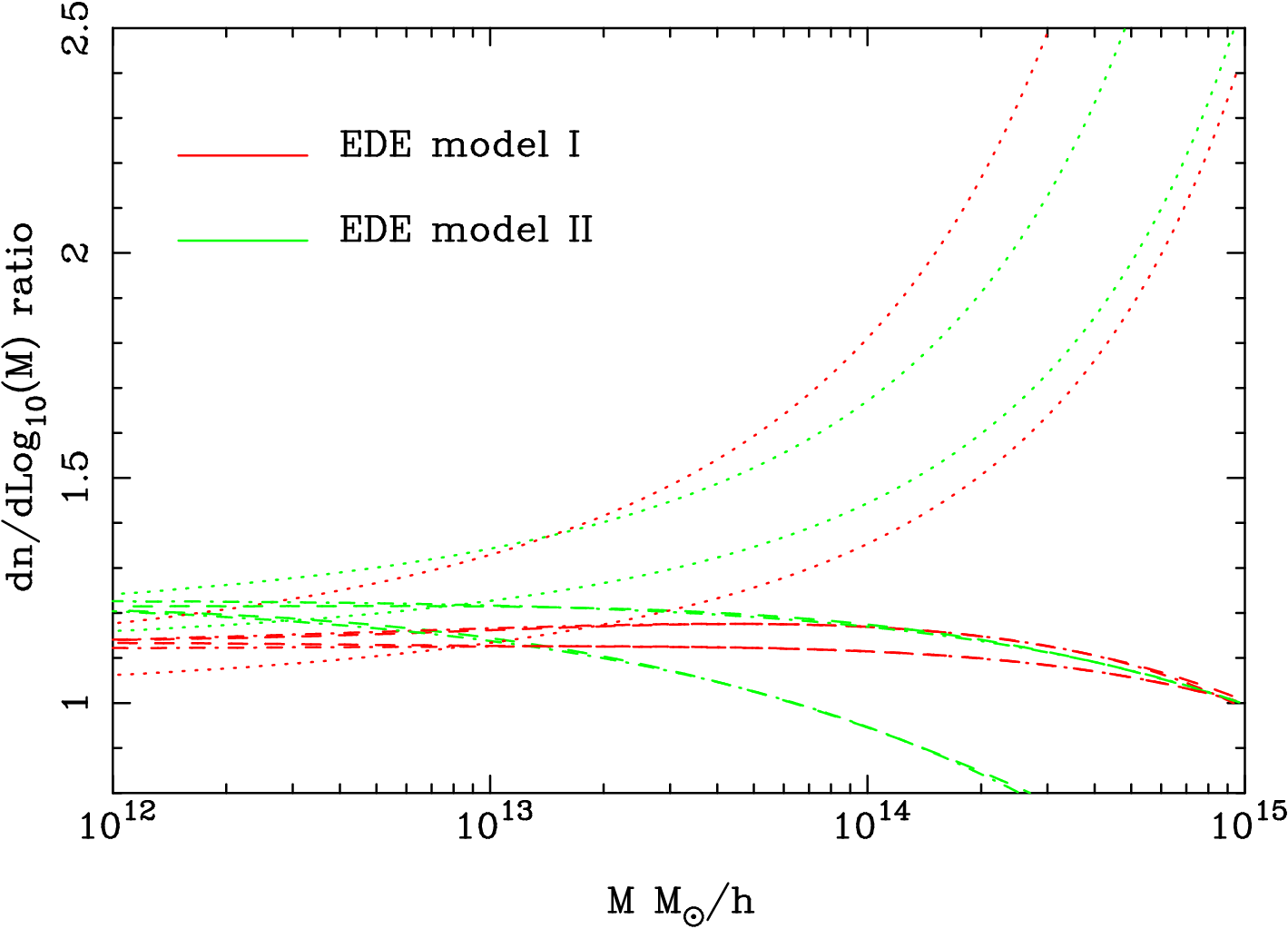}
  \caption{Reproduction  of  Fig.  5  of \citet{bartDW06}  with  the
  \citet{Jenkins01}    and   \citet{Warren}   mass    functions   also
  included. The Jenkins and Warren  formulas are shown with dashed and
  dot-dashed lines respectively, however they are indistinguishable on
  this  scale.   They  both  clearly disagree  with  the  Sheth-Tormen
  prediction,  shown with  dotted  lines, for  both  models.  For  the
  Sheth-Tormen mass function,  the upper curves are for  $z=1$ and the
  lower  $z=0$,  however for  the  Jenkins-Warren  mass functions  the
  reverse is the case, the lower curves are for $z=1$.}\label{bartfig}
\end{figure}

However, in this work we compare the two approaches for EDE and find a
marked  qualitative  difference  in nonlinear  structure  predictions,
although  they both  use the  same linear  density behavior.   We have
reproduced Fig.  5  of \citet{bartDW06} which showed the  ratio of the
HMF  to the  $\Lambda$CDM  case for  two  EDE models  (both using  the
bending    parametrisation    of    \citet{Wetterich04},    see    our
Eq. \ref{wetEDE}),  at $z=0$ and $z=1$.  However,  in our reproduction
(see Fig.   \ref{bartfig}) we have also included  the prediction using
instead  the  Jenkins and  Warren  formulas  (note  that these  curves
overlap  and can  barely be  distinguished on  the scale  shown).  The
Sheth-Tormen mass functions shown  utilised the method for calculating
the  linear density  contrast at  collapse time,  $\delta_c$,  for EDE
cosmologies from \citet{bartDW06}, which  results in a quite different
value  from $\Lambda$CDM.  As  can  be seen,  the  predictions of  the
Sheth-Tormen and the Jenkins-Warren approaches are very different, and
it is  not clear {\it a priori}  which approach should be  used in the
absence of rigorous calculation via N-body simulations.  In this work,
we fill the gap in  the literature by performing these simulations and
halo analyses.  Note that the  difference in the halo abundance ratios
predicted by  the Jenkins  and Warren formulas  is negligible  for all
cosmologies studied in  this work and in subsequent  figures we plot a
single  line   that  represents  the  prediction  of   both  of  these
approaches.

\subsection{Press-Schechter vs. \ Jenkins-Warren}\label{bartmods}

As a  direct comparison  to the work  of \citet{bartDW06},  we perform
simulations  of the  same two  EDE  and one  $\Lambda$CDM models  they
considered   analytically.   Model   I   has  $\Omega_{m,0}h^2=0.146,$
$\Omega_bh^2=0.026,$     $h=0.67,$    $n=1.05,$     $w_0=-0.93$    and
$\sigma_8=0.82$;     Model     II     has     $\Omega_{m,0}h^2=0.140,$
$\Omega_bh^2=0.023,$    $h=0.62,$     $n=0.99,$    $w_0=-0.99,$    and
$\sigma_8=0.78$.   Both models  have an  averaged value  for  the dark
energy density during the matter dominated era of $\Omega_{d,sf}=0.04$
(see \citet{omsf} for the precise definition).

\begin{figure}
\includegraphics[
  scale=0.35,
  angle=-90]{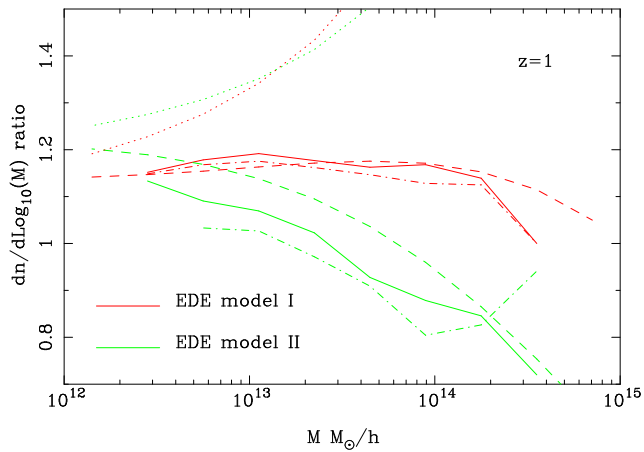}
  \caption{Mass function ratios at  $z=1$ between EDE and $\Lambda$CDM
 for  different methods  of predicting  the mass  function.  The solid
 lines  show  the  simulation  data  with halos  found  with  the  FOF
 method. The dot-dashed lines show  the SO method results for the same
 data. The  dashed lines show  the Jenkins mass function  (which gives
 almost  the same  prediction  as the  Warren  formula). Finally,  the
 dotted lines  show the prediction of the  Sheth-Tormen mass function.
 The  Jenkins-Warren formula  is clearly  preferred by  the simulation
 data.}\label{DMF2280}
\end{figure}

\begin{figure}
\includegraphics[
  scale=0.35,
  angle=-90]{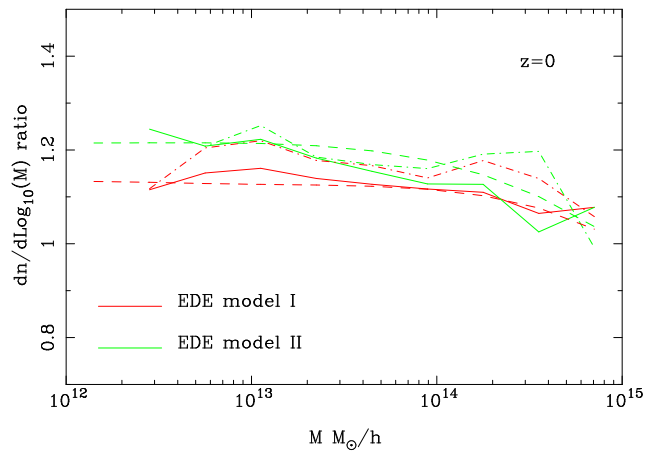}
  \caption{Mass function  ratios at $z=0$ for the  models described in
  Section  \ref{bartmods}.  The  line styles  are the  same  as Fig.
  \ref{DMF2280}.}\label{DMF2281}
\end{figure}

The  simulation  results for  the  mass  function,  overlaid with  the
predictions  from  both   the  Sheth-Tormen  and  Jenkins-Warren  mass
functions, are shown in  Figs. \ref{DMF2280} and \ref{DMF2281}.  The
simulations show a good agreement  with the predictions of the Jenkins
formula and  strong disagreement  with the Sheth-Tormen  function with
the appropriate  EDE linear collapse parameter.  This  has two crucial
implications.  The difference between  $\Lambda$CDM and EDE in numbers
of collapsed  objects is much less than  previously thought. Secondly,
the abundance increase reverses  at higher redshift, i.e.  rather than
an order of  magnitude {\it more} halos in EDE  at $z=1$ as previously
predicted, we  find {\it less} halos at  the high mass end  for one of
the models, and comparable number of halos at the low mass end.

This result is important and troubling; instead of the number of halos
being exponentially sensitive and a crucial tool for detecting EDE, it
appears that  it is quite insensitive  to the existence of  EDE at the
few  percent  energy  density  level,  at or  below  the  upper  limit
satisfying current  data.  The  implications are further  discussed in
Section \ref{conclude}.

\subsection{Early Dark Energy Mass Functions}\label{haloresults}

Although we  used the  bending model of  EDE for direct  comparison to
\citet{bartDW06}, an  improved model of EDE was  formulated after that
paper.  We treat the EDE  model of \cite{DoranRobbers06}, given in Eq.
\ref{edeparam}, as the standard description of EDE in the remainder of
the  paper.  It  is  somewhat  more closely  related  to the  particle
physics  models and automatically  goes to  a constant  energy density
contribution  $\Omega_e$  at  early  times  in  both  the  matter  and
radiation eras.

\begin{figure}
\includegraphics[
  scale=0.35,
  angle=-90]{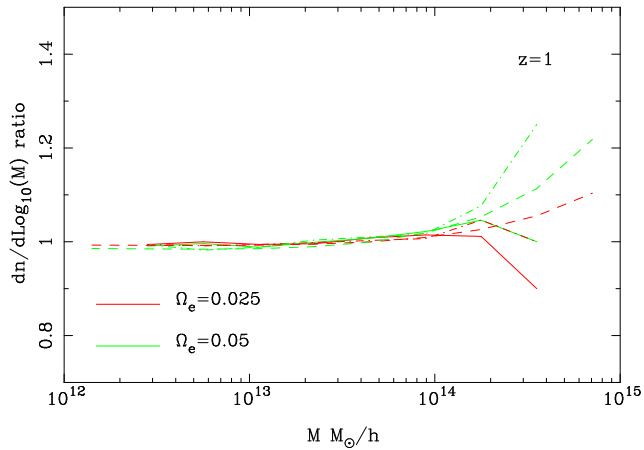}
  \caption{Mass function ratios relative  to $\Lambda$CDM at $z=1$ for
  two EDE  models of  the form in  Eq.  \ref{edeparam}  with different
  values of  $\Omega_e$ but with the cosmology  fixed otherwise. The
  dashed lines  show the  Jenkins-Warren mass function  prediction for
  the ratios while  the solid lines show the  simulation results using
  the  FOF method and  the dot-dashed  line the  results using  the SO
  method.}\label{DMF2171}
\end{figure}

\begin{figure}
\includegraphics[
  scale=0.35,
  angle=-90]{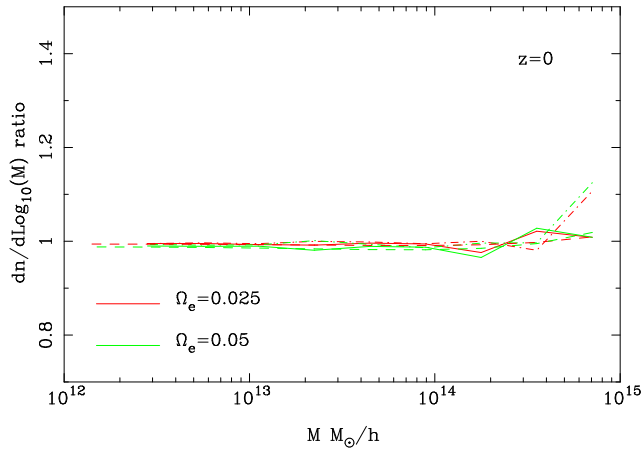}
  \caption{Mass function ratios at $z=0$. The line styles are the same
  as Fig. \ref{DMF2171}.}\label{DMF2172}
\end{figure}

We  compare  the  simulation  results  for this  EDE  model  with  the
predictions  of the Jenkins-Warren  formulas in  Figs. \ref{DMF2171}
and  \ref{DMF2172}.   Note  that  we  do  not  plot  the  Sheth-Tormen
predictions for the mass function, since we find that using the linear
collapse   parameter  derivation   from   \citet{bartDW06}  with   the
parametrisation of  \citet{2005JCAP...10..011D} leads to  a limit that
does not converge  \footnote{If one assumes that the  overdensity of a
spherical  perturbation, $\Delta-1$,  is  proportional to  $a$ (as  in
\citet{bartDW06}), then since in  the Doran-Robbers model $D_+(a) \sim
a^{1-(3\Omega_e)/5}$,   then   their   ratio   in   Eq.    5   becomes
$a^{(3\Omega_e/5)}$  and so $\delta_c$  formally goes  to zero  in the
limit.   To calculate the  collapse parameter  one actually  needs the
full  numerical  calculation.}   and  hence  we  cannot  determine  an
appropriate  $\delta_c$.  In  any case,  the results  of  the previous
section  indicated  that  this  approach  was  not  suitable  for  EDE
cosmologies.

As in  Section \ref{bartmods}  there are two  key points to  note, the
first  is  that  the   difference  between  $\Lambda$CDM  and  EDE  is
negligible  at  $z=0$ and  at  $z=1$  (see  Figs. \ref{DMF2171}  and
\ref{DMF2172}),  and  second that  the  Jenkins-Warren mass  functions
correctly predict this  result.  Choosing to define the  halos via the
FOF or the SO method makes little difference to the ratio results.

The  simulations  used  in  this  study are  insufficient  to  make  a
precision determination of the accuracy of the Jenkins-Warren formulas
for  EDE cosmologies,  however  it is  clear  that they  are at  least
accurate  to $\lesssim 10\%$  for predicting  the {\it  relative} mass
function of  EDE and  $\Lambda$CDM and future  work on EDE  should use
these rather than the linear theory collapse motivated mass functions.
If  required in  the future,  larger simulations  with a  greater mass
resolution  could be  used  to  determine the  mass  function for  EDE
cosmologies at percent level accuracy,  however this is outside of the
scope of this  work, and unnecessary at this stage due  to the lack of
precision in the corresponding observations.

\section{Power Spectrum}\label{psresults}

The  mass  power  spectrum   is  central  to  extracting  cosmological
information  from  galaxy  redshift  and  weak  lensing  surveys.   An
important  question  is  how  does  the addition  of  EDE  change  the
non-linear  mass power  spectrum? Ideally  we would  like to  be able,
through simulations,  to derive an  understanding of the  changes that
EDE makes  to the power  spectrum, in order  that we can  predict this
statistic  for   any  EDE   model  without  needing   further  lengthy
simulations.   The  section  presents  the  results  of  a  systematic
variation  of  cosmological parameters,  in  order  to understand  the
non-linear power spectrum in EDE cosmologies.

The approach we take is to examine how the ratio of the non-linear EDE
power  spectrum  and  $\Lambda$CDM  compares  to  {\it  linear}  power
spectrum ratios. In  order to turn this information  into a prediction
of  the EDE  non-linear power  spectrum in  general  requires accurate
prior knowledge of the absolute value of the non-linear power spectrum
for $\Lambda$CDM,  for instance the simulation  calibrated, halo model
motivated, formula from \citet{Halofit} (know as Halofit), or any more
accurate future  method.  In  addition, EDE changes  the shape  of the
linear CDM  power spectrum.  We  therefore must calculate  the correct
linear   power   spectrum,   using   a   package   such   as   CMBEASY
\citep{2005JCAP...10..011D}, as used in  this study.  Armed with these
two methods, we  could then predict the non-linear  power spectrum for
EDE  models {\it  if we  could relate  the ratio  between the  EDE and
$\Lambda$CDM  linear and  non-linear power  spectra}.  These  will not
generally be the same.  This section addresses the ratio relation.

Determining  the  absolute value  of  the  power  spectrum is  a  more
intensive numerical task, outside the scope of this work.  However, as
argued in \citet{2006MNRAS.366..547M} and \citet{2007MNRAS.380.1079F},
many numerical  errors will  cancel in  a ratio, and  hence we  can be
confident,  as  long  as  convergence  is  demonstrated  (see  Section
\ref{simdetail}) that the ratio  results are robust.  Another issue is
that  the presence  of  EDE shifts  the  sound horizon  and hence  the
locations  in  $k$  space of  the  BAO  peaks  in the  power  spectrum
\citep{DST07,LinderRobbers}.  However,  our simulations concentrate on
the broad  features of the power  spectrum and do  not have sufficient
volume or statistics to resolve the percent level shift.  Furthermore,
the problem  of how subtle non-linear  effects alter the  BAO peaks in
general   is  an   ongoing  area   of  research   (see   for  instance
\citet{2008MNRAS.383..755A}). The focus of  our current analysis is on
the  comparison between cosmologies  of the  broad features  and small
scales.

A  simple  first  step  in  predicting the  non-linear  power  in  EDE
cosmologies might  be to use the  Halofit formula with  the EDE linear
power and  growth factor. In  the subsequent sections, we  present the
predictions of  this modified Halofit  form along with  our simulation
results to  see how useful such an  approach may be. In  order to make
this  calculation  the  effective  spectral index  $n_{\rm  eff}$  and
curvature $C$ were determined from the EDE linear power spectrum.  See
\citet{Halofit} for details and the Halofit formula.

\subsection{Varying $\Omega_e$}\label{varyoe}

\begin{figure}
\includegraphics[
  scale=0.35,
  angle=-90]{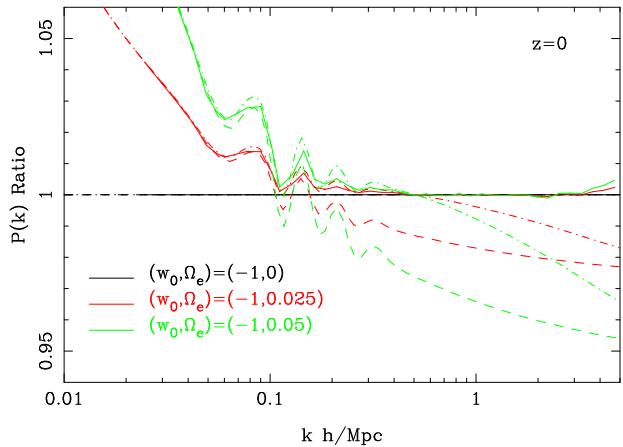}
  \caption{Keeping  $\sigma_{8}$  constant  while  adding  early  dark
  energy  results  in  a   remarkably  similar  non-linear  power  for
  $k\gtrsim0.2\,h$/Mpc.  This is despite  the ratios in the linear power
  spectra, shown in the dashed lines, diverging.  The modified Halofit
  prediction is shown by the dot-dashed lines.}\label{s8matcha1}
\end{figure}

\begin{figure}
\includegraphics[
  scale=0.35,
  angle=-90]{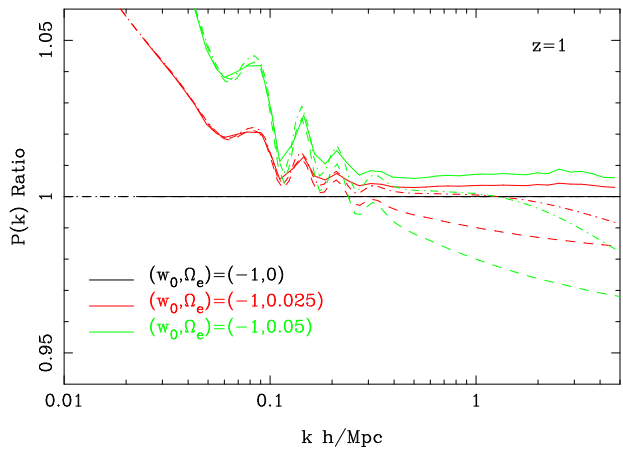}
  \caption{At  $z=1$  models  with  early dark  energy  have  slightly
  increased non-linear power on small scales, however the shape of the
  relative   non  linear   power   curve  differs   from  the   linear
  power.}\label{s8matcha0.5}
\end{figure}

We  start with  a standard  concordance flat  $\Lambda$CDM  model with
parameters as given in  Section \ref{simdetail}.  We then consider two
EDE  models  with  different  amounts  of dark  energy  in  the  early
universe,  $\Omega_e$,  while  keeping  the  rest  of  the  cosmology,
including the growth  normalisation today, $\sigma_8$, unaltered.  The
linear and  non-linear mass power  spectrum ratios at $z=0$  and $z=1$
are shown in Fig.  \ref{s8matcha1} and \ref{s8matcha0.5}.

As discussed in Section  \ref{simdetail}, the numerical convergence of
the  power spectrum  ratios has  been  established at  the $\sim  1\%$
level. The dependence of the ratio on numerical factors (primarily box
size  and particle  resolution)  is generally  greater  at higher  $k$
values.  The results past $k\gtrsim3\,h$/Mpc should perhaps be treated
with more  caution, but at lower  $k$ values the  simulations are well
converged.

Compared to  $\Lambda$CDM, for a fixed primordial  spectral index $n$,
the presence of  dark energy in the early  universe changes the higher
$k$ slope  of the post recombination linear  power spectrum, resulting
in more large  scale, but less small scale, linear  power for the same
overall normalisation  $\sigma_8$.  The growth amplitude  in the early
universe must be higher in  EDE models than $\Lambda$CDM to compensate
for the lower growth rate, in  order to get the same $\sigma_8$ today.
The  relative  non-linear  power  between EDE  and  $\Lambda$CDM  will
therefore depend  on those two  governing effects, i.e. the  change in
the linear  growth rate history and  the change in  the initial linear
power spectrum shape.  Non-linear power  on small scales is coupled to
the power  on large scales  and this non-linear amplification  will be
enhanced in the EDE models relative to $\Lambda$CDM due to the greater
amplitude  of  large  scale  power  (see  Figs.   \ref{s8matcha1}  and
\ref{s8matcha0.5}).  The  Halofit formula prediction  (dot-dashed line
in Figs.  \ref{s8matcha1} and \ref{s8matcha0.5}) should be expected to
take  into  account at  least  some  of  the non-linear  amplification
effects  for the  EDE linear  power spectrum,  since this  formula was
calibrated based  on a variety  of spectral shapes. While  the correct
linear growth factor  at any given redshift can  be calculated for EDE
and  used  in the  Halofit  formula for  the  power  spectrum at  that
redshift, the modified growth {\it  history} of EDE models relative to
$\Lambda$CDM,     will     not      be     taken     into     account.
\citet{2007MNRAS.380.1079F} and \citet{2007ApJ...665..887M} found that
the linear growth history in the early universe is an important factor
in determining the non-linear growth  at later epochs where the linear
growth are  matched.  Therefore the  scales where Halofit  breaks down
should  grant an  insight  into  the relative  importance  of the  two
factors, non-linear amplification and  the linear growth history, as a
function of scale.

\begin{figure}
\includegraphics[
  scale=0.35,
  angle=-90]{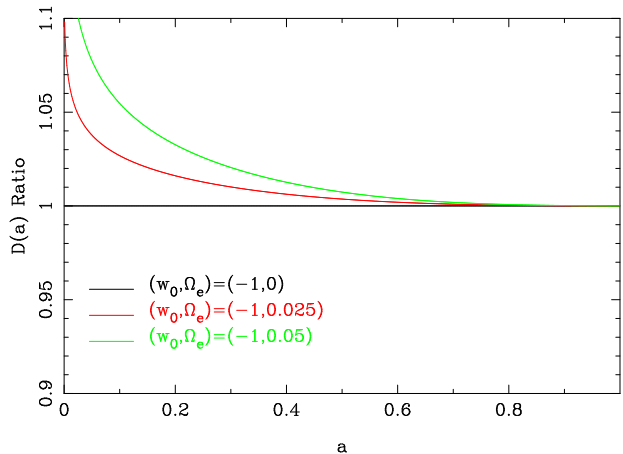}
  \caption{The  linear growth  factors  $D(a)/D(1)$ of  models
  with  early  dark  energy   relative  to  $\Lambda$CDM.  The  linear
  evolution of the  growth is identical to within  ~1\% after $z\simeq
  1$.}\label{gfac201}
\end{figure}

The results show  that the prediction of the  modified Halofit formula
reproduces the  simulation results  in the translinear  regime, around
$k=0.3-1\,h$/Mpc.  As  expected, the greater amplitude  of large scale
power in the EDE models leads to enhanced non-linear amplification and
hence the non-linear  power ratio is greater than  the linear.  Beyond
$k\gtrsim1\,h$/Mpc, Halofit under-predicts  the non-linear power, with
the difference between this  and simulations increasing with $k$. This
can  be understood  from the  ratio in  linear growth  history  of the
models  shown   in  Fig.  \ref{gfac201}.   The   smallest  scales  are
remembering the conditions of expansion, and linear growth, of earlier
epochs, leading to  a greater relative amount of  non-linear growth at
those scales.

\begin{figure}
\includegraphics[
  scale=0.35,
  angle=-90]{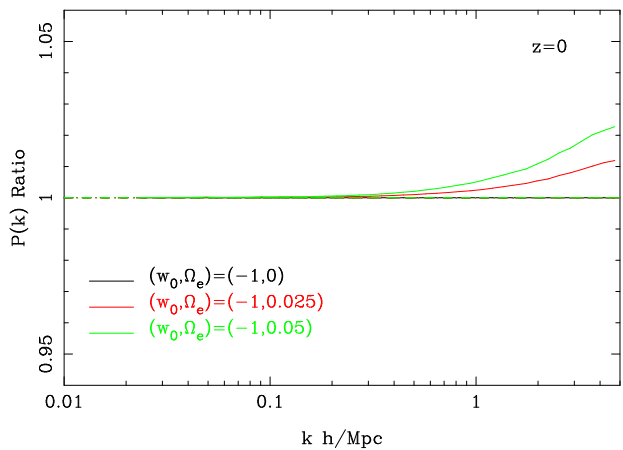}
  \caption{The same  background cosmologies as  Fig. \ref{s8matcha1}
  but  using   the  $\Lambda$CDM   initial  power  spectrum   for  all
  models. The higher amplitude of  growth in the early universe in the
  EDE  models leads  to a  greater  non-linear power  at small  scales
  today.}\label{commonPk}
\end{figure}

In order  to decouple the effects  of the modified  growth history and
spectral shape, we performed  simulations simulations in which the EDE
models used the $\Lambda$CDM  initial power spectrum while maintaining
their  correct   expansion  history.   The  $z=0$   result  for  these
simulations are  shown in Fig.  \ref{commonPk}.  In  this case Halofit
predicts no  difference between the  models, due to the  common linear
power   spectrum    shape   and   normalisation.     Examining   Figs.
\ref{s8matcha1} and \ref{commonPk} it  can be seen that the difference
between the  simulations and Halofit  predictions in both  figures are
comparable.  This reinforces the  idea that Halofit correctly predicts
much of the  effects of the EDE spectral  shape on intermediate scales
and the modified  growth history increases the power  at small scales.

This comparison is not perfect and  we should not expect to be able to
completely decompose  the effects of spectral shape  and linear growth
history  into  a  linear  sum.  These  results  compare  to  those  of
\citet{modGrav}  (see their  Fig. 8)  for  the effects  of a  modified
gravitational  potential on  the non-linear  power spectrum.   In that
case,  the   effects  on  the  non-linear  growth   of  modifying  the
gravitational potential were found  to be negligible compared with the
altered linear power spectrum shape.

Note  however  that this  is  essentially  the expectation  previously
proposed  from analytic  predictions of  the halo  mass  function. The
higher amplitude  of growth  in the early  Universe in EDE  models was
expected to produce more collapsed structures at high redshift, with a
residual increase in  collapsed structures at low redshift  due to the
greater  formation rate  in the  early Universe.   Our  power spectrum
results do  indicate an enhanced amount of  non-linear growth relative
to linear  growth for the EDE  models, however this  enhancement is at
the percent level rather than  the considerable increase that would be
seen if the relative halo abundance were an order of magnitude at high
redshift as previously predicted.

Interestingly, the  differences in the non-linear growth  rates due to
the spectral  shape and  linear growth history  appear to  conspire to
wash out  the early dark energy signal  at $z=0$. That is  to say, the
non-linear  power in  the EDE  and $\Lambda$CDM  models is  matched to
within  a percent  for scales  smaller than  $k>0.2\,h$/Mpc.   Even at
$z=1$ (Fig. \ref{s8matcha0.5}) the high $k$ power in the EDE models is
well   matched,  to  about   $(\Omega_e/0.05)$\%.   Since   we  expect
$\Omega_e\lesssim 0.03$  from current observational  constraints, this
is  a remarkable  match. This  result accords  with the  mass function
results   of   Section   \ref{haloresults}  that   found   essentially
indistinguishable abundances  of dark matter  halos. Note that  we are
holding  the cosmology  apart from  $\Omega_e$ fixed  between  EDE and
$\Lambda$CDM.   More generally we  would not  expect this  match.  For
instance, a steep primordial spectral  index $n$ in an EDE model would
compensate for the  later flattening of the linear  power spectrum due
to  the presence  of  EDE leading  to  a linear  power spectrum  after
recombination much  more similar to  $\Lambda$CDM.  In this  case, as
shown  in Fig.  \ref{commonPk}, we  would expect  a greater  amount of
non-linear power in the EDE model.

\subsection{Varying the Linear Growth History}

\begin{figure}
\includegraphics[
  scale=0.35,
  angle=-90]{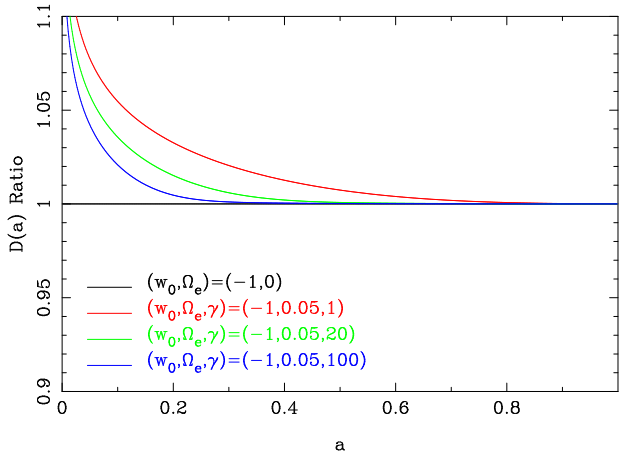}
  \caption{The effect  of varying  the $\gamma$ parameter  in Eq.
  \ref{extendedEDEparam} on the  linear growth factor.  Increasing the
  value shifts the departure  from the fiducial $\Lambda$CDM to higher
  redshifts.}\label{vgammaGrowth}
\end{figure}

We now  examine in more detail  the influence of the  linear growth at
different epochs  on the non-linear power.   In bottom-up hierarchical
growth, the  smallest scales reach the non-linear  growth rate earlier
than larger  scales.  We therefore  expect a relationship  between the
relative non-linear  growth at different  scales imprinted on  the low
redshift power spectrum and the  alteration of the growth rate at high
redshifts. This relationship was found to be important for scales with
$k\gtrsim1\,h$/Mpc in Section  \ref{varyoe}. To investigate this further,
we  have employed the  extended EDE  parametrisation suggested  in the
appendix of \citet{DoranRobbers06}

\begin{equation}\label{extendedEDEparam}
\Omega_d(a)              =              \frac{\Omega_d^0             -
\Omega_e(1-a^{-3w_0})^\gamma}{\Omega_d^0+\Omega_m^0a^{3w_0}}        +
\Omega_e(1-a^{-3w_0})^\gamma
\end{equation}
where $\gamma$ controls the importance  of the early dark energy terms
at  late times.  Increasing  $\gamma$ pushes  to higher  redshifts the
epoch  when this model  changes from  dark energy  with a  with nearly
constant equation of state  $w=w_0$ (i.e.\ $\Lambda$CDM when $w_0=-1$)
to one with appreciable early  dark energy density (or $w$ approaching
0). This  extended parametrisation reduces to  Eq. \ref{edeparam} when
$\gamma=1$.  In \citet{DoranRobbers06} it was found that including the
extra  parameter   does  not  alter  the  constraints   on  the  other
cosmological  parameters  when  fitted  to  the data  and  hence  this
additional  parameter  was   not  deemed  necessary.   However,  this
parameter is  useful in the current  study as an  aid to understanding
the dependence of the relative  non-linear growth as a function of the
high  redshift linear  growth behaviour.   The effect  of  varying the
$\gamma$  parameter on  the linear  growth factor  is shown  in Fig. \ref{vgammaGrowth}

\begin{figure}
\includegraphics[
  scale=0.35,
  angle=-90]{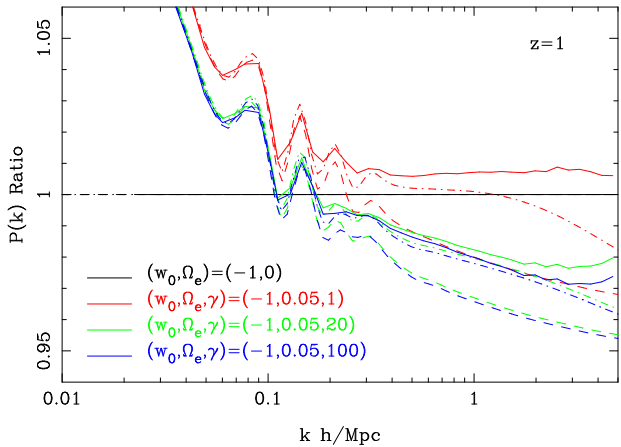}
  \caption{Power  spectrum ratios  with a  varying $\gamma$  factor at
  $z=1$. The  solid lines show  the non-linear power while  the dashed
  lines show  the linear  power, both as  a ratio to  the $\Lambda$CDM
  model.}\label{vgammaNLa0.5}
\end{figure}

\begin{figure}
\includegraphics[
  scale=0.35,
  angle=-90]{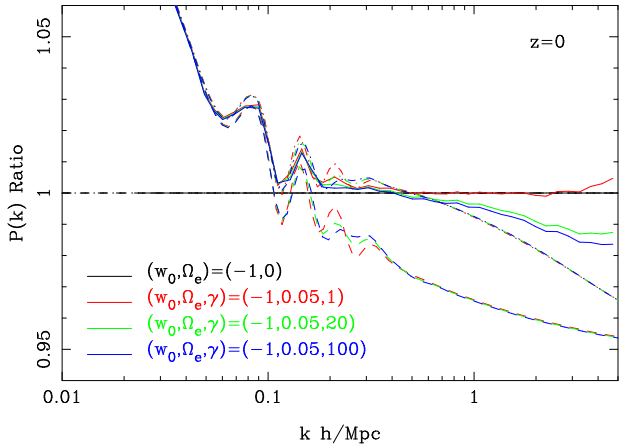}
  \caption{Power  spectrum ratios  with a  varying $\gamma$  factor at
  $z=0$. The  solid lines show  the non-linear power while  the dashed
  lines show  the linear  power, both as  a ratio to  the $\Lambda$CDM
  model.}\label{vgammaNLa1}
\end{figure}

How does  this altered linear  growth behaviour change  the non-linear
growth? The models shown  in Fig. \ref{vgammaGrowth} were simulated,
with the  measured non-linear power spectrum results  shown in Figs.
\ref{vgammaNLa0.5} and \ref{vgammaNLa1}.

At  $z=0$  (Fig.  \ref{vgammaNLa1}),  models with  a  higher  $\gamma$
parameter have  less non-linear power  at small scales than  the model
with  $\gamma=1$. This  can  be understood  from  the relative  linear
growth      histories      of      these     models      shown      in
Fig. \ref{vgammaGrowth}.  Since the  high $\gamma$ models  remain more
like the fiducial $w=-1$ to higher redshift, the accumulated non-linear
effects due to the linear growth history are smaller in this case than
for $\gamma=1$. This causes the non-linear power to more closely track
the Halofit prediction.

At  $z=1$  (Fig. \ref{vgammaNLa0.5})  Halofit  matches the  simulation
results well  for the  high $\gamma$ models.  Note that for  both high
$\gamma$  models, the  linear growth  factor evolution  is essentially
identical to $\Lambda$CDM  in the period between $z=1$  and $z=0$ (see
Fig. \ref{vgammaGrowth}).  Despite this, the non-linear ratio of these
models  to $\Lambda$CDM  does change  between these  two  epochs. This
increase is  therefore due to the  extra large scale power  in the EDE
models, and is predicted by Halofit.

Altering  the linear  growth  history through  the $\gamma$  parameter
demonstrates the relationship between  the high redshift linear growth
history and  the low redshift small  scale power. As  found in Section
\ref{varyoe}, this relationship, while important, is not strong enough
to cause a significant increase in small scale power in EDE models. The
small scale power at low and intermediate redshift is sensitive to the
high redshift linear growth history at only the percent level.

\subsection{Varying the Equation of State}

\begin{figure}
\includegraphics[
  scale=0.35,
  angle=-90]{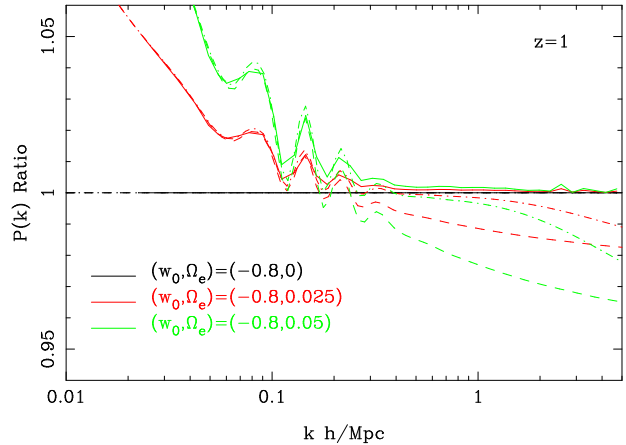}
  \caption{Power  spectrum ratios  relative to  a  fiducial $w_0=-0.8$
  constant  equation  of  state  model.   The  solid  lines  show  the
  non-linear   power  while   the   dashed  lines   show  the   linear
  power.   Compared  to  Fig.  \ref{s8matcha0.5},  there   is  less
  difference  between   the  linear  and   non-linear  power  spectrum
  ratios. }\label{vwaNLa0.5}
\end{figure}

Until now  we have compared EDE  models to $\Lambda$CDM  only. We will
now  investigate the  effects from  a different  fiducial  dark energy
equation  of state  today,  $w_0$.  We  ran  a series  of models  with
$w_0=-0.8$,  with  the   rest  of  the  cosmology  the   same  as  the
$\Lambda$CDM  model  from Section  \ref{varyoe}.   The relative  power
results showed very little dependence on $w_0$, with the $z=0$ results
almost  indistinguishable   from  those  with   $w_0=-1$.   At  higher
redshifts there is a weak  dependence on $w_0$ in the non-linear power
ratio,   as   seen  by   comparing  the   high  $k$   plateaus  of
Fig. \ref{vwaNLa0.5}  for $w_0=-0.8$  vs.\  Fig. \ref{s8matcha0.5}  for
$w_0=-1$.

In Fig. \ref{vwaNLa0.5}, the ratios of the linear and non-linear power
spectra  are  more  similar  than  in  Fig.   \ref{s8matcha0.5}  where
$w_0=-1$. This can be understood from the linear growth history which,
compared  to Fig.  \ref{gfac201}, shows  a reduced  difference  in the
linear growth history of the  models at high redshift.  This is simply
due  to the $w_0=-0.8$  case dark  energy acting  more like  matter in
general  (due  to the  less  negative  equation  of state)  hence  the
addition  of  the  EDE  component  has  less  influence  than  in  the
$\Lambda$CDM case.

\subsection{Varying the Growth Amplitude}\label{varys8}

\begin{figure}
\includegraphics[
  scale=0.35,
  angle=-90]{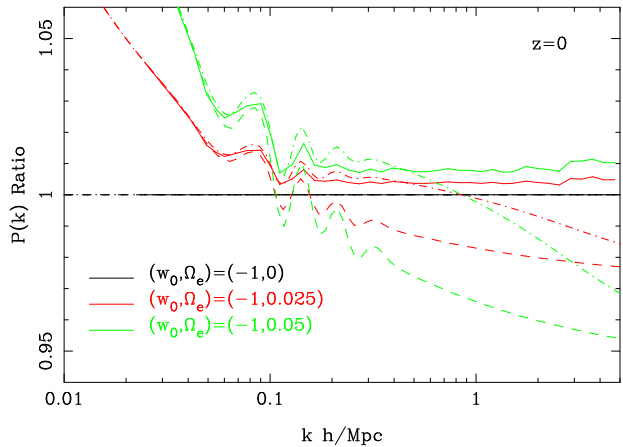}
  \caption{Power  spectrum  ratios  relative  to $\Lambda$CDM  with  a
  higher $\sigma_8$,  0.96, than that for  Fig. \ref{s8matcha1} with
  $\sigma_8=0.76$.   With a  greater amplitude  of  growth, non-linear
  effects  are  enhanced  and  the non-linear  power  spectrum  ratios
  diverge more from the linear ratios.}\label{s8varya1}
\end{figure}

The overall amplitude  of structure is a major  factor determining the
rate of non-linear growth. It is therefore important to understand how
sensitive  the  relative  non-linear  growth  is  to  this.   We  have
performed a series of simulations varying the linear theory $\sigma_8$
defined  at $z=0$.   The result  for $\sigma_8=0.96$  (remembering the
original model shown in  Fig.  \ref{s8matcha1} had $\sigma_8=0.76$) at
$z=0$ is shown  in Fig.  \ref{s8varya1}.  As might  be expected, while
$\sigma_8$ scales out in the  linear power ratios (so the dashed lines
are the  same as  in Fig.  \ref{s8matcha1}),  the higher  amplitude of
growth leads  to enhanced non-linear  effects.  The difference  in the
non-linear  and  linear  power   spectrum  ratios  is  increased  with
increasing $\sigma_8$.  The results  for $z=1$ have a similar increase
in the power spectrum ratio and are not shown for brevity.

Once again  we can see that  the modified Halofit  formula matches the
simulation results well at  intermediate scales but under predicts the
power in  the EDE  models on smaller  scales. The scale  where Halofit
begins to under predict the EDE  power is pushed to a lower $k$ value,
again due to  the increased non-linear effects from  the higher growth
amplitude normalisation.

\begin{figure}
\includegraphics[
  scale=0.35,
  angle=-90]{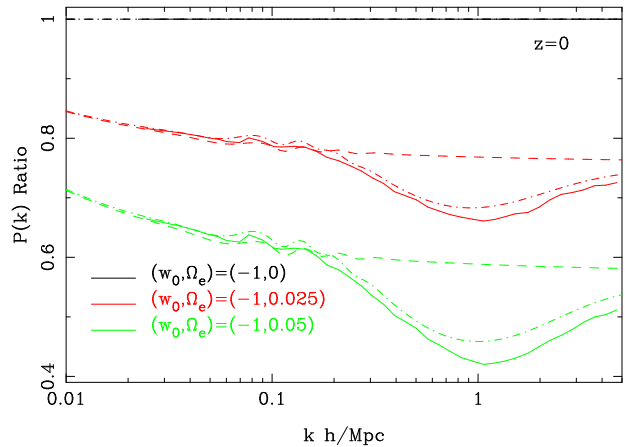}
  \caption{Simulations  with fixed  primordial amplitude  $A_s$ rather
  than fixed $\sigma_8$ today.}\label{asfix}
\end{figure}

To  test our  intuition  about effects  governing  the power  spectrum
results, we also ran  simulations with fixed primordial power spectrum
amplitude $A_s$ rather  than $\sigma_8$. The $z=0$ result  is shown in
Fig. \ref{asfix}.   For these  simulations, one gets  different linear
power  today (and  so different  $\sigma_8$) but  the early  epochs of
collapse, showing up in the  very nonlinear (higher $k$) regime today,
are more  similar. This can be  seen in the power  for $k>1\,h$/Mpc in
the  EDE models  which increases  with $k$,  heading back  towards the
$\Lambda$CDM  power since  the  amplitude of  power  in the  different
models was more similar when those scales collapsed than it is today.

Matching $\sigma_8$ appears to be  sufficient to ensure that the small
scale non-linear power between $\Lambda$CDM and EDE models are matched
to  within a  few  percent, regardless  of  the amount  of early  dark
energy, $\Omega_e$  (see Fig. \ref{s8matcha1}) or  the redshift when
this EDE  term turns on  (see Fig.  \ref{vgammaNLa1}).  This  match is
weakly dependent  on $\sigma_8$; however varying $\sigma_8$  by 0.2, a
considerable amount, shifts the ratio in power by only a percent.

These results leads  to a useful tool for  confronting EDE models with
power spectrum  measurements, such as galaxy redshift  surveys or weak
lensing. The Halofit  formula, modified by using the  EDE linear power
spectrum  and growth factor,  predicts the  non-linear power  ratio to
within a  percent out to  $k\sim1\,h$/Mpc.  For smaller  scales, using
the $\Lambda$CDM non-linear  power matches the EDE result  to within a
few  percent or  so,  particularly  at low  and  zero redshift.   This
prescription is  fairly loose but could  be improved upon  with a more
quantitative  study  of  the  variation  in  power  with  cosmological
parameters.  The  main point  from this study  is that  the variations
between EDE and $\Lambda$CDM are at the few percent level only, rather
than there being a significant difference in non-linear power.

\subsection{Growth vs.\ Geometry}

\begin{figure}
\begin{center}
\includegraphics[
  scale=0.35,
  angle=0]{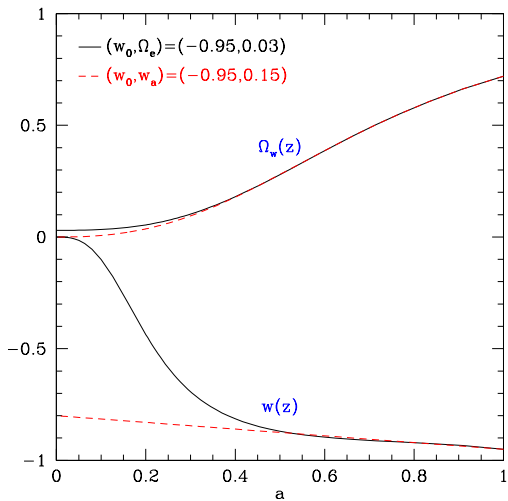}
  \caption{Dark energy density and equation of state comparison for an
  EDE and  $(w_0,w_a)$ model with  distance history matched  to 0.02\%
  out to $z=2$.}\label{EL1}
\end{center}
\end{figure}

\begin{figure}
\includegraphics[
  scale=0.35,
  angle=-90]{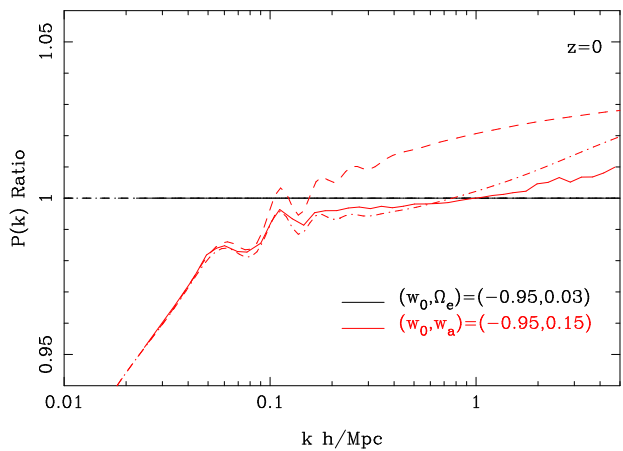}
  \caption{Ratio results  at $z=0$ for the comparison  between the two
  parameterisations for  dark energy given  by $w(a)=w_0+w_a(1-a)$ and
  Eq.  \ref{edeparam}, where the  parameter values are chosen match to
  the distance history.}\label{dmatcha1}
\end{figure}

\begin{figure}
\includegraphics[
  scale=0.35,
  angle=-90]{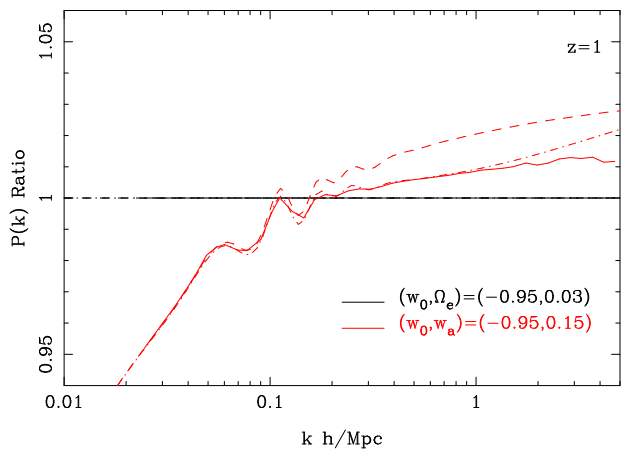}
  \caption{As Fig. \ref{dmatcha1} but at $z=1$.}\label{dmatcha0.5}
\end{figure}

The expansion  history of an  EDE model can  be fit well out  to $z=2$
with  a  time  varying  equation  of  state  $w(a)=w_0+w_a(1-a)$  with
$w_a\approx  5\Omega_e$.   Fig. \ref{EL1}  shows  the  dark  energy
density  and equation  of  state for  an  EDE model  and the  matching
$(w_0,w_a)$  model.   Distance  measurements,  such as  from  Type  Ia
supernovae,  will agree  in the  two models  to 0.02\%  out  to $z=2$.
However, the  early universe histories were very  different.  Does the
growth history  distinguish between these two  models?  To investigate
this  question,  we  simulated  a   pair  of  models,  one  using  the
Eq. \ref{edeparam} form of  EDE with $\Omega_e=0.03$ and $w_0=-0.95$
and one using the $(w_0,w_a)$ form of dark energy with $w_0=-0.95$ and
$w_a=0.15$ with  the cosmology matched otherwise.   The power spectrum
ratio    results   are   shown    in   Figs.   \ref{dmatcha1}   and
\ref{dmatcha0.5}.

\begin{figure}
\includegraphics[
  scale=0.35,
  angle=-90]{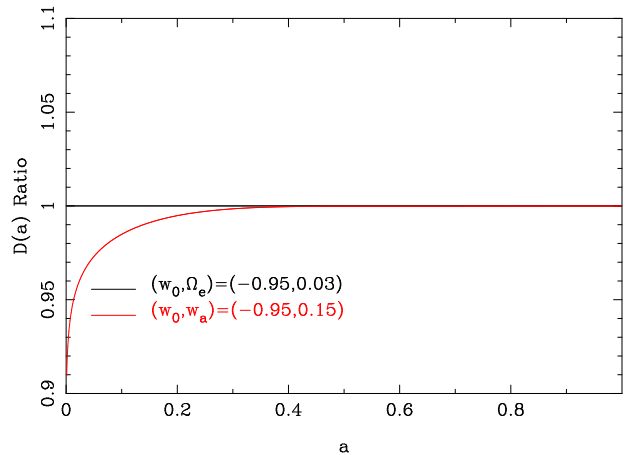}
  \caption{The relative linear growth factor evolution for the EDE and
 $(w_0,w_a)$ models with matching distance history.}\label{dmatchG}
\end{figure}

In this  case the EDE model is  taken as the fiducial,  and we display
the ratio  of the $(w_0,w_a)$ model  to this. Once again,  we see that
the non-linear power  on small scales is well  matched between the two
models, despite the linear power  ratios not matching at these scales.
At $z=1$  the difference between  the linear and non-linear  ratios is
less, with  more non-linear power  at small scales in  the $(w_0,w_a)$
model.   Once again  it is  instructive to  examine the  linear growth
history, shown in Fig. \ref{dmatchG}.

This figure shows that the EDE  model has a higher amplitude of growth
in the early  universe than the $(w_0,w_a)$ model. This  is due to the
greater amount  of dark energy present  at this time,  leading to less
linear growth  over the whole  history of the universe,  requiring the
EDE model to start at a higher growth amplitude in order to obtain the
same linear theory $\sigma_8$ today.  The linear growth factor is very
well  matched between  the two  models after  $z\approx  2$ ($a\approx
0.33$), yet  as can be seen,  the non-linear growth  between $z=1$ and
$z=0$ is  different in the two  models, again due  to the nonlinearity
amplification effect.

\section{Conclusions}\label{conclude}

In this work we have examined  the effects of early dark energy on the
non-linear growth  of structure  through N-body simulations.   The key
result  is  that  non-linear  large  scale  structure,  including  the
abundance of  clusters, is relatively  insensitive to the  presence of
EDE.   This result  is in  contrast  to previous  analytic work  which
expected a  substantial increase  in non-linear structure  compared to
$\Lambda$CDM with the same $\sigma_8$.

The implications  of the results presented  here are that  EDE is much
more  difficult to detect  than previously  thought.  This  presents a
problem  for  dark  energy  cosmology,   not  only  for  the  sake  of
understanding dark  energy but because, as shown  in \citet{DST07} and
\citet{LinderRobbers},  the  presence of  EDE  can significantly  bias
distance  measurements  using  baryon  acoustic  oscillations  if  the
possibility  of  EDE  is  not  considered.  Fitting  for  EDE  however
significantly reduces  the discriminating  power of BAOs.  

Ideally it  was hoped  to detect the  presence of EDE  through cluster
abundances   (see   \citet{FedeliBart07}),   such   as   with   future
Sunyaev-Zel'dovich,  weak  lensing, X-ray,  or  optical surveys.   This
would  provide  an  independent  measure  of EDE  thus  restoring  the
discriminating  power   of  BAO  measurements.    However,  this  work
demonstrates that this is not the case.

As the  amount of EDE increases,  it becomes easier  to distinguish in
that the primordial amplitude  of matter density perturbations must be
increased  to  reach the  same  growth  ($\sigma_8$)  by today.   Such
effects in  linear growth and  the CMB anisotropy power  spectra allow
that EDE must  contribute less than a few percent  to the early energy
density  \citep{DoranRW07}.   The  acoustic  sound horizon  shifts  by
approximately  half   of  this.   However  this   article  shows  that
observations of non-linear structure should not be expected to provide
a substantial leap in our ability to detect EDE.

This study  has demonstrated that  the halo mass function  formulas of
both \citet{Jenkins01} and \citet{Warren} are valid for predicting the
EDE to $\Lambda$CDM  results ratio to within $\sim  10\%$ or better in
the mass  range $\sim 10^{12}  - 10^{15} M_\odot$.  Future  studies of
EDE should  use these formulas or N-body  simulations when considering
halo  abundances,  rather   than  spherical  collapse  motivated  mass
functions as have been used in the past.

We have also demonstrated that the Halofit \citep{Halofit} formula for
the non-linear power spectrum, when  using the EDE linear matter power
spectrum and growth factor, is accurate for EDE cosmologies for scales
larger than  around $k\sim1\,h$/Mpc.   On smaller scales  this formula
under predicts the  power. Further work  would be needed  to accurately
re-calibrate this formula  for EDE on these scales,  perhaps along the
lines  of the  way \citet{2006MNRAS.366..547M}  corrected  Halofit for
constant $w$ dark energy cosmologies.  Until such a study is performed
we propose, as discussed in  Section \ref{varys8}, that the results of
this study suggest a simple prescription for predicting the non-linear
power spectrum for EDE models to within an accuracy of a percent or so
relative to the $\Lambda$CDM prediction.

\section*{Acknowledgments}

We thank  the Centre for Astrophysics and  Supercomputing at Swinburne
University for  access to the  supercomputing facilities used  in this
project. Georg Robbers and Michael Doran are thanked for providing the
early  dark energy  implementation  for CMBEASY.   Alexander Knebe  is
thanked for  helpful discussions about AHF.   GFL acknowledges support
from ARC Discovery Project DP0665574.  This work has been supported in
part  by  the Director,  Office  of  Science,  Office of  High  Energy
Physics,  of  the  U.S.\  Department  of Energy  under  Contract  No.\
DE-AC02-05CH11231.

\bsp

\label{lastpage}

\end{document}